# Teleparallel Superspace in Eleven Dimensions Coupled to Supermembranes [1]

S. James Gates, Jr.,[†]   Hitoshi Nishino[⋆]   and   Subhash Rajpoot[#]

[†]  *Dept. of Phys., Univ. of Maryland, College Park, MD 20742-4111*
gatess@wam.umd.edu

[⋆] *Dept. of Phys. & Astr., California State Univ., Long Beach, CA 90840*
hnishino@csulb.edu

[#]  *Dept. of Phys. & Astr., California State Univ., Long Beach, CA 90840*
rajpoot@csulb.edu

## Abstract

We present a superspace formulation of $N=1$ eleven-dimensional supergravity with no manifest local Lorentz covariance, which we call teleparallel superspace. This formulation will be of great importance, when we deal with other supergravity theories in dimensions higher than eleven dimensions, or a possible formulation of noncommutative supergravity. As an illustrative example, we apply our teleparallel superspace formulation to the case of $N=1$ supergravity in twelve-dimensions. We also show the advantage of teleparallel superspace as backgrounds for supermembrane action.

---

[1]This work is supported in part by NSF grant # PHY-93-41926.

# 1. Introduction

Local Lorentz covariance is usually taken for granted in the conventional supergravity theories [1][2], starting with the $N=1$ supergravity in four dimensions (4D)[2] [3], up to the $N=1$ supergravity in 11D [4][5]. However, the recent developments in higher dimensional supergravity in $D \geq 12$ [6], or of an $N=2$ supergravity theory [7][8] motivate formulations in which the local Lorentz symmetry is not manifest. As a matter of fact, in the consideration of generalized algebra called M-algebra [9] for M-theory [10], the manifest Lorentz covariance is not crucial, but is realized only as hidden symmetry in terms of composite connections. We call such superspace, in which no local Lorentz symmetry is manifest, 'teleparallel superspace'.[3]

In conventional supergravity theories, there has been no strong motivation of teleparallelism formulation. This is because in component formulation of conventional supergravity, the local Lorentz covariance is manifest from the outset anyway, and moreover, there is no strong reason to consider its explicit breakings[4], because of the need of local Lorentz covariance for removing unwanted ghosts in the system. This situation in component turns out to be similar in superspace formulations [11].[5] As such, there has been so far no spectacular development for teleparallel supergravity, until the recent construction of higher-dimensional supergravity in $D \geq 12$ with no manifest Lorentz covariance [6][7][8]. Even though these higher-dimensional supergravity theories have no built-in local Lorentz covariance, they have become important nowadays, motivated also by M-theory [10] and F-theory [12], or S-theory [13], and so forth.

Another important motivation of considering the lack of manifest Lorentz covariance is associated with the recent development of noncommutative geometry, in which the introduction of 'constant tensor' $\theta^{\mu\nu}$ explicitly breaks Lorentz covariance [14]. For example, in a recent paper [15] we have shown that teleparallel gravity can be the foundation of noncommutative gravity, in order to delete the undesirable negative energy ghosts in the antisymmetric component $B_{\mu\nu}$ in the complex metric $g_{\mu\nu}$ in the Lorentz covariant formulation [16]. From the viewpoint that the existence of the constant tensor $\theta^{\mu\nu}$ already breaks the Lorentz symmetry, it is even imperative to consider the supersymmetric extension, *i.e.*, teleparallel supergravity.

In this paper we present a teleparallel superspace formulation with a consistent set of superspace constraints for Bianchi identities, together with the physical superfield equations. We also couple this background to supermembrane action, and confirm its fermionic $\kappa$-symmetry. Interestingly, we find that the fermionic $\kappa$-symmetry becomes even simpler and more natural in the teleparallel superspace formulation. Such a formulation will be of great importance when we deal with higher dimensional supergravity [6][7][8] in which the local Lorentz symmetry is not manifest. As an illustrative example, we describe how to reformulate $N=1$ superspace supergravity in 12D [6][8] in teleparallel superspace.

---

[2]We also use the notation $D=4$ such as in $D=4, N=1$ supergravity.

[3]We adopt this terminology for superspace, in which local Lorentz covariance is not manifest, even if it is *not* explicitly broken. In this sense, our terminology covers a wider set of formulations in superspace. This formulation is also similar to the superspace for $D=4, N=1$ supergravity in [11].

[4]See, *e.g.*, subsection 1.5 of ref. [1].

[5]Some superspace formulation in 4D in this direction was presented in [11], using only the torsion superfield with no curvature superfields, in order to reduce possible counter-terms for renormalizability of supergravity.

## 2. Notational Preliminaries

Before giving our constraints, we first set up our notations, which are conceptually different from the conventional supergravity. Our superspace covariant derivative $E_A$ is simply defined by

$$E_A \equiv E_A{}^M \partial_M \quad , \tag{2.1}$$

with no term for Lorentz connection $\phi_{Ab}{}^c$ [2]. Relevantly, the anholonomy coefficients $C_{AB}{}^C$ are defined by

$$C_{AB}{}^C \equiv (E_{[A} E_{B)}{}^M) E_M{}^C \quad , \tag{2.2}$$

with no explicit Lorentz connection here, either. Similarly, the superfield strength $F_{ABCD}$ is defined by

$$F_{ABCD} \equiv \tfrac{1}{6} E_{[A} A_{BCD)} - \tfrac{1}{4} C_{[AB|}{}^E A_{E|CD)} \quad . \tag{2.3}$$

with no Lorentz connection. Accordingly, the superspace Bianchi identities are only $C$- and $F$-types:

$$E_{[A} C_{BC)}{}^D - C_{[AB|}{}^E C_{E|C)}{}^D \equiv 0 \quad , \tag{2.4}$$

$$\tfrac{1}{24} E_{[A} F_{BCDE)} - \tfrac{1}{12} C_{[AB|}{}^F F_{F|CDE)} \equiv 0 \quad . \tag{2.5}$$

We sometimes refer them respectively as $(ABC, D)$- and $(ABCDE)$-type Bianchi identities. Even though we do not give a detailed computation, the l.h.s. of both (2.4) and (2.5) are covariant under the following local Lorentz transformation with the parameter $\Lambda^{ab}$ [2], despite the absence of Lorentz connection $\phi_{Ab}{}^c$:

$$\delta_{\mathrm{L}} E_A{}^M = \tfrac{1}{2} \Lambda^{bc} (\mathcal{M}_{bc})_A{}^B E_B{}^M \quad , \tag{2.6a}$$

$$\delta_{\mathrm{L}} C_{AB}{}^C = +\tfrac{1}{2} \Lambda^{bc} (\mathcal{M}_{bc})_{[A|}{}^D C_{D|B)}{}^C + \tfrac{1}{2}(-1)^{(A+B)(C+D)} \Lambda^{bc} (\mathcal{M}_{bc})^{CD} C_{ABD}$$
$$\qquad + \tfrac{1}{2}(E_{[A|} \Lambda^{bc})(\mathcal{M}_{bc})_{|B)}{}^C \quad , \tag{2.6b}$$

$$\delta_{\mathrm{L}} F_{ABCD} = \tfrac{1}{12} \Lambda^{ab} (\mathcal{M}_{ab})_{[A|}{}^E F_{E|BCD)} \quad , \tag{2.6c}$$

where (2.6b) follows from (2.6a). The local Lorentz covariance of the l.h.s. of (2.4) and (2.5) is confirmed by the mutual cancellations of terms generated by the local Lorentz transformation (2.6). To put it differently, the local Lorentz symmetry is a 'hidden' symmetry in this formulation.

Accordingly, the Ricci tensor component $R_{ab}(\phi)$ in the conventional superspace can be re-expressed in terms of the anholonomy coefficients by

$$R_{ab}(\phi) = - E_a C_b - \tfrac{1}{2} E_c C_{ab}{}^c - \tfrac{1}{2} E_c C^c{}_{(ab)} - C_{ac}{}^\gamma C_{\gamma b}{}^c$$
$$+ \tfrac{1}{2} C_a{}^{cd} C_{bdc} + \tfrac{1}{2} C_a{}^{cd} C_{bcd} + \tfrac{1}{2} C_{ab}{}^c C_c + \tfrac{1}{2} C^c{}_{(ab)} C_c - \tfrac{1}{4} C_{cda} C^{cd}{}_b \ , \qquad (2.7)$$

Here $C_a \equiv C_{ab}{}^b$ is the vectorial component of $C_{ab}{}^c$. Note that (2.7) is an equality only to understand the total local Lorentz covariance, because the l.h.s. makes sense only in the conventional superspace with the explicit Lorentz connection, while the r.h.s. in terms of the anholonomy coefficients is useful only in our teleparallel superspace. In other words, we should not use such expressions as the l.h.s. of (2.7) in our teleparallelism. Interestingly, the antisymmetric component $R_{[ab]}$ is shown to vanish identically due to the $(abc, c)$-type Bianchi identity:

$$R_{[ab]} = -E_{[a} C_{b]} - E_c C_{ab}{}^c + C_{ab}{}^c C_c - C_{[a|c}{}^\gamma C_{\gamma|b]}{}^c \equiv 0 \ , \qquad (2.8)$$

therefore (2.6) has effectively only the symmetric component in $R_{(ab)}$.

## 3. Constraints and Superfield Equations

There are basically two methods to fix superspace constraints for our teleparallel superspace. The first method is to use the relationships between the torsion superfields in the conventional superspace and the teleparallel superspace, and another way is the direct way, writing down all the possible term with unknown coefficients to be fixed by the satisfactions of all the Bianchi identities (2.1) and (2.2). Both of these methods give the following consistent set of constraints:

$$C_{\alpha\beta}{}^c = +i(\gamma^c)_{\alpha\beta} \ , \qquad (3.1a)$$

$$F_{\alpha\beta cd} = +\tfrac{1}{2}(\gamma_{cd})_{\alpha\beta} \ , \qquad (3.1b)$$

$$C_{\alpha\beta}{}^\gamma = +\tfrac{1}{4}(\gamma_{de})_{(\alpha}{}^\gamma C_{\beta)}{}^{de} \ , \qquad C_\alpha{}^{bc} = -C_\alpha{}^{cb} \ , \qquad (3.1c)$$

$$C_{ab}{}^\gamma = +\tfrac{i}{144}(\gamma_b{}^{[4]} F_{[4]} + 8\gamma^{[3]} F_{b[3]})_\alpha{}^\gamma - \tfrac{1}{8}(\gamma^{cd})_\alpha{}^\gamma (2C_{bcd} - C_{cdb}) \ , \qquad (3.1d)$$

$$E_\alpha C_{\beta cd} = +\tfrac{1}{144}(\gamma_{cd}{}^{[4]})_{\alpha\beta} F_{[4]} + \tfrac{1}{6}(\gamma^{[2]})_{\alpha\beta} F_{cd[2]} - \tfrac{i}{4}(\gamma^e)_{\alpha\beta} C_{cde} + \tfrac{i}{4}(\gamma^e)_{\alpha\beta} C_{e[cd]}$$
$$+ \tfrac{1}{4}(\gamma^{ab})_\alpha{}^\gamma C_{\beta ab} C_{\gamma cd} + C_{\alpha c}{}^e C_{\beta ed} \ , \qquad (3.1e)$$

$$E_\alpha C_{bcd} = +E_{[b|}C_{\alpha|c]d} + C_{\alpha[b|}{}^\epsilon C_{\epsilon|c]d} + C_{\alpha[b|}{}^e C_{e|c]d} - i(\gamma_d C_{bc})_\alpha + C_{bc}{}^e C_{e\alpha d} \ , \tag{3.1f}$$

$$E_\alpha F_{bcde} = -\tfrac{1}{8}(\gamma_{[bc}C_{de]})_\alpha + \tfrac{1}{6}C_{\alpha[b|}{}^f F_{f|cde]} \ , \tag{3.1g}$$

$$E_\gamma C_{ab}{}^\delta = +E_{[a|}C_{\gamma|b]}{}^\delta + C_{ab}{}^e C_{e\gamma}{}^\delta + C_{ab}{}^\epsilon C_{\epsilon\gamma}{}^\delta + C_{\gamma[a|}{}^e C_{e|b]}{}^\delta + C_{\gamma[a|}{}^\epsilon C_{\epsilon|b]}{}^\delta \ . \tag{3.1h}$$

We sometimes use the symbol $F_{[4]}$ for quantities with totally antisymmetric bosonic indices to save space, namely $F_{[4]}$ is equivalent to $F_{abcd}$. Eq. (3.1a) through (3.1e) satisfy the Bianchi identities of the engineering dimensions $d = 0$, $1/2$ and $d = 2$, while (3.1f) through (3.1h) are from $d = 3/2$ and $d = 2$.

Let now us give some remarks about these constraints. Note first that even though the particular component $C_{ab}{}^c$ remains in many of these constraints, it is recombined to form a locally Lorentz covariant expression in the physical superfield equations to be seen later. As was mentioned, even though there is the second direct method to obtain these constraints, it is easier to use the first method using the relationships between the conventional superspace and our teleparallel superspace. For example, a set of superspace constraints for torsion superfields in the former [5][17] are expressed in terms of anholonomy coefficients as

$$T_{\alpha\beta}{}^c = C_{\alpha\beta}{}^c \ , \tag{3.2a}$$

$$T_{abc} = C_{abc} - \phi_{[ab]c} = 0 \quad \Longleftrightarrow \quad \phi_{abc} = \tfrac{1}{2}(C_{abc} - C_{acb} - C_{bca}) \ , \tag{3.2b}$$

$$T_{\alpha b}{}^c = C_{\alpha b}{}^c - \phi_{\alpha b}{}^c = 0 \quad \Longrightarrow \quad C_{\alpha b}{}^c = \phi_{\alpha b}{}^c \ , \quad \phi_{\alpha(bc)} = C_{\alpha(bc)} = 0 \ , \tag{3.2c}$$

$$T_{\alpha\beta}{}^\gamma = C_{\alpha\beta}{}^\gamma + \tfrac{1}{4}(\gamma_{de})_{(\alpha|}{}^\gamma \phi_{|\beta)}{}^{de} = 0 \quad \Longrightarrow \quad C_{\alpha\beta}{}^\gamma = -\tfrac{1}{4}(\gamma_{de})_{(\alpha|}{}^\gamma \phi_{|\beta)}{}^{de} \ , \tag{3.2d}$$

$$T_{\alpha b}{}^\gamma = C_{\alpha b}{}^\gamma - \phi_{b\alpha}{}^\gamma = C_{\alpha b}{}^\gamma + \tfrac{1}{4}(\gamma^{cd})_\alpha{}^\gamma \phi_{bcd} \ , \tag{3.2e}$$

$$T_{ab}{}^\gamma = C_{ab}{}^\gamma \ . \tag{3.2f}$$

For example, from (3.11) we see that we can impose the condition that the symmetric component in $C_\alpha{}^{cd}$ is zero as in (3.1c). Therefore if we adopt essentially the same constraints for the conventional superspace as in [5][17], we see that $C_{\alpha\beta}{}^c$ is exactly the same as $T_{\alpha\beta}{}^c$ as in (3.1) from (3.2), $\phi_{abc}$ is completely expressed in terms of $C_{abc}$ as in (3.10) as usual, $C_{\alpha b}{}^\gamma$ is expressed in terms of $F_{[4]}$ and $C_{abc}$ as in (3.1e) via (3.2f), $C_{\alpha\beta}{}^\gamma$ is expressed in terms of $C_{\alpha b}{}^c$, while $C_{ab}{}^\gamma$ can be regarded as the superfield strength in our superspace, equivalent to $T_{ab}{}^\gamma$ in the conventional superspace.

Once we comprehend these correspondences, the satisfaction of all the Bianchi identities (2.4) and (2.5) is clear. However, we can also confirm all the Bianchi identities by the usual direct computation.

Prepared with these constraints, we can also get the superfield equations from the $d = 3/2$ and $d = 2$ Bianchi identities, as

$$i(\gamma^b)_{\alpha\beta} C_{ab}{}^\beta = 0 \ , \tag{3.3a}$$

$$E_a C_b + \tfrac{1}{2} E_c C_{ab}{}^c + E_d C^d{}_{(ab)} - \tfrac{1}{2} C_a{}^{cd} C_{bdc} - \tfrac{1}{2} C_{acd} C_b{}^{cd} - \tfrac{1}{2} C_{ab}{}^c C_c$$
$$- \tfrac{1}{2} C^c{}_{(ab)} C_c + \tfrac{1}{4} C_{cda} C^{cd}{}_b + C_{ac}{}^\gamma C_{\gamma b}{}^c$$
$$+ \tfrac{1}{3} F_{a[3]} F_b{}^{[3]} - \tfrac{1}{36} \eta_{ab} F_{[4]}{}^2 = 0 \ , \tag{3.3b}$$

$$E_d F_{abc}{}^d - \tfrac{1}{4} C_{de[a} F_{bc]}{}^{de} - \tfrac{1}{2} C^d F_{abcd} + \tfrac{1}{576} \epsilon_{abc}{}^{[4][4]'} F_{[4]} F_{[4]'} = 0 \ . \tag{3.3c}$$

Eqs. (3.3a), (3.3b) and (3.3c) are respectively the gravitino, gravitational and $F$-superfield equations. The $EC$, $C^2$, and $C_{ac}{}^\gamma C_{\gamma b}{}^c$-terms[6] in the first two lines in (3.3b) are all arranged themselves to be equivalent to the Ricci tensor (2.7) in the conventional superspace, and therefore the total expression (3.3b) is local Lorentz covariant. This explains also why the $C_{ac}{}^\gamma C_{\gamma b}{}^c$-term is needed in there. As usual in any 11D superspace formulation, the gravitino superfield equation (3.15) is obtained by the $(\alpha\beta\gamma, \delta)$-Bianchi identity at $d = 3/2$, while the gravitational superfield equation (3.3b) is obtained by the spinorial derivative $(\gamma^c)^{\gamma\alpha} E_\gamma [ i(\gamma^a)_{\alpha\beta} C_{ab}{}^\beta ] = 0$, and similarly the $F$-field equation (3.3c) is from $(\gamma_{[ab|})_\alpha{}^\beta E_\beta [ i(\gamma^d)^\alpha{}_\gamma C_{|c]d}{}^\gamma ] = 0$. The last two are also based on the constraints (3.1g) and (3.1h). Recall that all the terms antisymmetric in [ab] in (3.3b) vanish identically due to (2.8).

## 4. Couplings to Supermembranes

Once our superspace has been established, our next natural step is to consider its couplings to supermembrane, in particular with fermionic symmetries.

Our total action for supermembrane theory is the same as that in the conventional one:

$$I \equiv I_\sigma + I_A \ , \tag{4.1}$$

$$I_\sigma \equiv \int d^3\sigma \left( +\tfrac{1}{2} \sqrt{-g} g^{ij} \eta_{ab} \Pi_i{}^a \Pi_j{}^b - \tfrac{1}{2} \sqrt{-g} \right) \ , \tag{4.2}$$

$$I_A \equiv \int d^3\sigma \left( -\tfrac{1}{3} \epsilon^{ijk} \Pi_i{}^A \Pi_j{}^B \Pi_k{}^C A_{CBA} \right) \ . \tag{4.3}$$

As usual, we have $\Pi_i{}^A \equiv (\partial_i Z^M) E_M{}^A$, and the indices $i, j, \cdots = 0, 1, 2$ are for the 3D world-volume for supermembrane.

In the conventional formulation, we need to vary $\Pi_i{}^A$ under the general transformations $\delta E^A \equiv (\delta Z^M) E_M{}^A$. In our teleparallel superspace, this variation is given by

$$\delta \Pi_i{}^A = \partial_i (\delta E^A) - \Pi_i{}^B (\delta E^C) C_{CB}{}^A \ . \tag{4.4}$$

Note that there is no Lorentz connection $\phi_{Ab}{}^c$ explicitly involved in here, as is usually the case with our teleparallel superspace. On the other hand, in the conventional superspace

---

[6]The $C$'s in $EC$ and $C^2$ are for $C_{ab}{}^c$ with purely bosonic indices.

[18][17] we have

$$\delta\Pi_i{}^A = \nabla_i(\delta E^A) - \Pi_i{}^B(\delta E^C)(T_{CB}{}^A + \phi_{CB}{}^A) \ , \tag{4.5}$$

where $\nabla_i(\delta E^A) \equiv \partial_i(\delta E^A) + (1/2)\Pi_i{}^B\phi_{Bc}{}^d(\mathcal{M}_d{}^c)^{AD}(\delta E_D)$, with the explicit Lorentz connection $\phi_{Ab}{}^c$. The explicit Lorentz connection term in (4.5), which looks in a sense unnatural, automatically decouples from the variation of the total action, due to $\delta_\kappa E^a = 0$ for the fermionic $\kappa$-transformation [18]. In our teleparallel superspace, on the other hand, the Lorentz connection term is absent from the outset, and there is no need for such a rearrangement.

Once this subtlety is clarified, the confirmation of fermionic $\kappa$-invariance [18] of our action: $\delta_\kappa I = 0$ under the transformation

$$\delta_\kappa E^\alpha = (I + \Gamma)^\alpha{}_\beta \kappa^\beta \ , \tag{4.6}$$

$$\delta_\kappa E^a = 0 \ , \tag{4.7}$$

$$\Gamma \equiv \frac{i}{6\sqrt{-g}} \epsilon^{ijk}\Pi_i{}^a\Pi_j{}^b\Pi_k{}^c(\gamma_{abc}) \ , \tag{4.8}$$

becomes straightforward. Here $\Gamma$ satisfies the following usual relationships under the embedding condition $g_{ij} = \Pi_i{}^a\Pi_{ja}$ equivalent to the $g_{ij}$-field equation:

$$\Gamma^2 = I \ , \quad \epsilon_i{}^{jk}\Pi_j{}^a\Pi_k{}^b\gamma_{ab}\Gamma = -2i\sqrt{-g}\,\Pi_i{}^a\gamma_a \ , \tag{4.9}$$

To summarize our supermembrane couplings, the fermionic $\kappa$-transformation becomes simpler and more natural in our teleparallel superspace, where there is no need of explicit Lorentz connection terms. We regard this feature as one of the advantages of considering teleparallelism in superspace formulation, as well as the first signal of the naturalness of teleparallel superspace for the couplings to supermembrane.

## 5. Supergravity in 12D as an Example

Once we have established the teleparallel formulation of superspace, the next natural step is to apply it to supergravity theories, in particular, to higher-dimensional supergravity in which Lorentz covariance is not built-in [6][7][8]. In this paper, we give the case of $N = 1$ supergravity in 12D [6][8] as an illustrative example.

We review first the *non*-teleparallel formulation of $N = 1$ supergravity in 12D [6][8] with important relationships. The most crucial ones is the definitions of the null-vectors $m^a$ and $n^a$:

$$(n^a) = \begin{pmatrix} (0) & (1) & \cdots & (9) & (11) & (12) \\ 0, & 0, & \cdots, & 0, & +\tfrac{1}{\sqrt{2}}, & -\tfrac{1}{\sqrt{2}} \end{pmatrix}, \quad (n_a) = \begin{pmatrix} (0) & (1) & \cdots & (9) & (11) & (12) \\ 0, & 0, & \cdots, & 0, & +\tfrac{1}{\sqrt{2}}, & +\tfrac{1}{\sqrt{2}} \end{pmatrix},$$

$$
(m^a) = \begin{pmatrix} \overset{(0)}{0}, & \overset{(1)}{0}, & \overset{\cdots}{\cdots}, & \overset{(9)}{0}, & \overset{(11)}{+\tfrac{1}{\sqrt{2}}}, & \overset{(12)}{+\tfrac{1}{\sqrt{2}}} \end{pmatrix}, \quad (m_a) = \begin{pmatrix} \overset{(0)}{0}, & \overset{(1)}{0}, & \overset{\cdots}{\cdots}, & \overset{(9)}{0}, & \overset{(11)}{+\tfrac{1}{\sqrt{2}}}, & \overset{(12)}{-\tfrac{1}{\sqrt{2}}} \end{pmatrix}.
$$
(5.1)

As in [6][8], we use the indices $a, b, \cdots = (0), (1), \cdots, (9), (10), (11)$ with parentheses for local Lorentz indices. Only in this section, we adopt the signature $(-, +, +, \cdots, +, -)$ in order to comply with the notation in [6][8]. We can also use the light-cone coordinates $V_\pm \equiv \tfrac{1}{\sqrt{2}}(V_{(11)} \pm V_{(12)})$. Relevantly, we have $n^a n_a = m^a m_a = 0$, $m^a n_a = m^+ n_+ = m^- n^- = +1$.[7] Other important quantities are the projection operators $P_\uparrow, P_\downarrow, P_{\uparrow\downarrow}$ defined by

$$P_\uparrow \equiv +\tfrac{1}{2}\not{n}\not{m} = +\tfrac{1}{2}\gamma^+\gamma^- \ , \qquad P_\downarrow \equiv +\tfrac{1}{2}\not{m}\not{n} = +\tfrac{1}{2}\gamma^-\gamma^+ \ , \tag{5.2a}$$

$$P_\uparrow P_\uparrow = +P_\uparrow \ , \quad P_\downarrow P_\downarrow = +P_\downarrow \ , \quad P_\uparrow + P_\downarrow = +I \ , \quad P_\uparrow P_\downarrow = P_\downarrow P_\uparrow = 0 \ , \tag{5.2b}$$

$$P_{\uparrow\downarrow} \equiv P_\uparrow - P_\downarrow = \gamma^{+-} \ . \tag{5.2c}$$

Relevantly, we have $(\not{n})_{\alpha\dot{\beta}} = -(\not{n})_{\dot{\beta}\alpha}$, $(\not{m})_{\alpha\dot{\beta}} = -(\not{m})_{\dot{\beta}\alpha}$, $(P_\uparrow)_{\alpha\beta} = -(P_\downarrow)_{\beta\alpha}$, $(P_{\uparrow\downarrow})_{\alpha\beta} = +(P_{\uparrow\downarrow})_{\beta\alpha}$. In supergravity in 12D, we can define the Lorentz generators formally as [6][8]

$$(\widetilde{\mathcal{M}}_{ab})^{cd} \equiv +\widetilde{\delta}_{[a}{}^c \widetilde{\delta}_{b]}{}^d \ , \tag{5.3a}$$

$$(\widetilde{\mathcal{M}}_{ab})_\alpha{}^\beta \equiv +\tfrac{1}{2}(\gamma_{ab} P_\uparrow)_\alpha{}^\beta \ , \qquad (\widetilde{\mathcal{M}}_{ab})_{\dot{\alpha}}{}^{\dot{\beta}} \equiv +\tfrac{1}{2}(P_\downarrow \gamma_{ab})_{\dot{\alpha}}{}^{\dot{\beta}} \ . \tag{5.3b}$$

where

$$\widetilde{\delta}_a{}^b \equiv \delta_a{}^b - m_a n^b = \begin{cases} \delta_i{}^j & \text{(for } a = i, \ b = j) \ , \\ \delta_+{}^+ = 1 & \text{(for } a = +, \ b = +) \ , \\ 0 & \text{(otherwise)} \ . \end{cases} \tag{5.4}$$

The Bianchi identities in the *non*-teleparallel superspace in 12D are [6][8]

$$\tfrac{1}{2}\nabla_{[A} T_{BC)}{}^D - \tfrac{1}{2} T_{[AB|}{}^E T_{E|C)}{}^D - \tfrac{1}{4} R_{[AB|e}{}^f (\widetilde{\mathcal{M}}_f{}^e)_{|C)}{}^D \equiv 0 \ , \tag{5.5}$$

$$\tfrac{1}{6}\nabla_{[A} G_{BCD)} - \tfrac{1}{4} T_{[AB|}{}^E G_{E|CD)} \equiv 0 \ , \tag{5.6}$$

$$\tfrac{1}{2}\nabla_{[A} R_{BC)d}{}^e - \tfrac{1}{2} T_{[AB|}{}^E R_{E|C)d}{}^e \equiv 0 \ . \tag{5.7}$$

One set of solutions for constraints at mass dimensions $0 \leq d \leq 1$ satisfying these Bianchi identities is [6][8]

$$T_{\alpha\beta}{}^c = +(\gamma^{cd})_{\alpha\beta} \nabla_d \varphi + (P_{\uparrow\downarrow})_{\alpha\beta} \nabla^c \varphi \ , \tag{5.8a}$$

$$G_{\alpha\beta c} = +T_{\alpha\beta c} \ , \tag{5.8b}$$

$$T_{\alpha\beta}{}^\gamma = +(P_\uparrow)_{(\alpha|}{}^\gamma (\gamma^c \overline{\chi})_{|\beta)} \nabla_c \varphi - (\gamma^{ab})_{\alpha\beta} (P_\downarrow \gamma_a \overline{\chi})^\gamma \nabla_b \varphi \ , \tag{5.8c}$$

$$\nabla_\alpha \Phi = +(\gamma^c \overline{\chi})_\alpha \nabla_c \varphi \ , \tag{5.8d}$$

---

[7]For other notational details, see refs. [6][8].

$$\nabla_\alpha \overline{\chi}_{\dot\beta} = -\tfrac{1}{24} \left(\gamma^{cde} P_\uparrow\right)_{\alpha\dot\beta} G_{cde} + \tfrac{1}{2} \left(\gamma^c P_\uparrow\right)_{\alpha\dot\beta} \nabla_c \Phi - (\gamma^c \overline{\chi})_\alpha \overline{\chi}_{\dot\beta} \nabla_c \varphi \ , \tag{5.8e}$$

$$G_{\alpha\beta\gamma} = 0 \ , \quad T_{\alpha b}{}^c = 0 \ , \quad T_{\alpha b}{}^\gamma = 0 \ , \quad G_{\alpha bc} = 0 \ , \tag{5.8f}$$

$$T_{ab}{}^c = -G_{ab}{}^c \ , \tag{5.8g}$$

where $\nabla_a \varphi$ and $\nabla_a \widetilde{\varphi}$ respectively satisfy the properties of $n_a$ and $m_a$ in (5.1), under the extra constraints

$$\nabla_{\underline{\alpha}} \varphi = \nabla_{\underline{\alpha}} \widetilde{\varphi} = 0 \ , \quad (\nabla_a \varphi)^2 = (\nabla_a \widetilde{\varphi})^2 = 0 \ , \quad (\nabla_a \varphi)(\nabla^a \widetilde{\varphi}) = 1 \ , \tag{5.9a}$$

$$\nabla_a \nabla_b \varphi = \nabla_a \nabla_b \widetilde{\varphi} = 0 \ . \tag{5.9b}$$

Here the underlined spinorial indices $\underline{\alpha}, \underline{\beta}, \cdots$ represent both dotted and undotted spinors: $\underline{\alpha} \equiv (\alpha, \dot\alpha), \underline{\beta} \equiv (\beta, \dot\beta), \cdots$ [6][8].

We have also extra constraints to delete *some* extra components of relevant superfields [6][8]:

$$T_{AB}{}^c \nabla_c \varphi = 0 \ , \quad G_{ABc} \nabla^c \varphi = 0 \ , \quad T_{aB}{}^C \nabla^a \varphi = 0 \ , \tag{5.10a}$$

$$R_{ABc}{}^d \nabla_d \varphi = 0 \ , \quad R_{aBc}{}^d \nabla^a \varphi = 0 \ , \tag{5.10b}$$

$$(\nabla^a \varphi) \nabla_a \Phi = 0 \ , \quad (\nabla^a \varphi) \nabla_a \overline{\chi}_{\dot\beta} = 0 \ , \tag{5.10c}$$

$$(\gamma^c)_\alpha{}^{\dot\beta} \overline{\chi}_{\dot\beta} \nabla_c \widetilde{\varphi} = 0 \ , \quad T_{ab}{}^\gamma (\gamma^d)_\gamma{}^{\dot\alpha} \nabla_d \widetilde{\varphi} = 0 \ , \tag{5.10d}$$

$$\phi_{Ab}{}^c \nabla_c \varphi = 0 \ , \quad \phi_{ab}{}^c \nabla^a \varphi = 0 \ . \tag{5.10e}$$

As has been stated in [6], these constraints will *not* delete *all* the extra components in 12D, and this is the non-trivial part of higher-dimensional supergravity [8].

We now consider the reformulation of teleparallel superspace of this $N=1$ supergravity in 12D [6][8]. In the *non*-teleparallel case, we had the supercurvature Bianchi identity (5.8) due to the existence of Lorentz connection $\phi_{Ab}{}^c$. Moreover, we had to confirm the consistency of this $R$-Bianchi identity (5.7) with other $T$ and $G$-Bianchi identities (5.5) and (5.6) [6]. However, in the present teleparallel superspace, we do not have the $R$-Bianchi identity. Instead we have only two Bianchi identities corresponding to (5.5) and (5.6):

$$E_{[A} C_{BC)}{}^D - C_{[AB|}{}^E C_{E|C)}{}^D \equiv 0 \ , \tag{5.11}$$

$$\tfrac{1}{6} E_{[A} G_{BCD)} - \tfrac{1}{4} C_{[AB|}{}^F G_{F|CD)} \equiv 0 \ . \tag{5.12}$$

The disappearance of the $R$-Bianchi identity is the first advantage for our teleparallel formulation for 12D supergravity, in which Lorentz covariance is not built-in off-shell from the outset. For example, even though the null-vectors $m_a$ and $n_a$ are replaced by more Lorentz 'covariant' gradients of scalar superfields in (5.9), these scalar superfields satisfy all the properties (5.1) only 'on-shell' as extra constraints, and therefore this formulation still has the drawback of lacking Lorentz covariance at the off-shell level. This gives a natural justification of considering teleparallel superspace formulation with no manifest local Lorentz covariance from the outset.

A set of constraints at mass dimensions $0 \leq d \leq 1$, satisfying the Bianchi identities (5.11) and (5.12) can be obtained in a way similar to (3.2). Namely, we first write down the relationships between the supertorsion $T_{AB}{}^C$, anholonomy coefficients $C_{AB}{}^C$, and Lorentz connection $\phi_{Ab}{}^c$, i.e., $T_{AB}{}^C \equiv C_{AB}{}^C + (1/2)\phi_{[A|d}{}^e(\widetilde{\mathcal{M}}_e{}^d)_{|B)}{}^C$, and next we solve this for $\phi_{Ab}{}^c$. Afterwards, we treat the anholonomy coefficients $C_{ab}{}^c$ and $C_{\alpha b}{}^c$ as independent superfields. The most important 'bridge' relationships between the *non*-teleparallel superspace and teleparallel superspace are

$$\phi_{\alpha bc} = C_{\alpha bc} + \phi_{\alpha[b|}{}^d (\nabla_{|c]}\varphi)(\nabla_d \widetilde{\varphi}) = -C_{\alpha cb} \quad , \tag{5.13a}$$

$$\phi_{abc} = +\tfrac{1}{2}(C_{a[bc]} - C_{bca} + G_{abc}) + \phi_{a[b|}{}^d(\nabla_{|c]}\varphi)(\nabla_d\widetilde{\varphi}) \quad , \tag{5.13b}$$

obtained from $T_{\alpha b}{}^c$ and $T_{ab}{}^c$ in (5.8f,g).

Compared with conventional superspace formulations [2], the last extra terms in these equations reflect the Lorentz non-covariance in our 12D. At first glance, these extra terms seem to cause a problem, we can not get closed forms for $\phi_{\alpha b}{}^c$ or $\phi_{ab}{}^c$. However, it does not matter in practice, because the extra terms in (5.13a,b) do *not* enter the combination $\phi_{Ad}{}^e(\widetilde{\mathcal{M}}_e{}^d)_B{}^C$ effectively, as can be easily confirmed under the definition of $\widetilde{\mathcal{M}}$ in (5.3).

After all of these, we get the set of constraints from (5.8) as

$$C_{\alpha\beta}{}^c = +(\gamma^{cd})_{\alpha\beta} E_d \varphi + (P_{\uparrow\downarrow})_{\alpha\beta} E^c \varphi \quad , \tag{5.14a}$$

$$G_{\alpha\beta c} = +C_{\alpha\beta c} \quad , \tag{5.14b}$$

$$C_{\alpha\beta}{}^\gamma = +(P_\uparrow)_{(\alpha|}{}^\gamma (\gamma^c \overline{\chi})_{|\beta)} E_c\varphi - (\gamma^{ab})_{\alpha\beta} (P_\downarrow \gamma_a \overline{\chi})^\gamma E_b\varphi + \tfrac{1}{4}(\gamma^{de} P_\uparrow)_{(\alpha|}{}^\gamma C_{|\beta)de} \quad , \tag{5.14c}$$

$$E_\alpha \Phi = +(\gamma^c \overline{\chi})_\alpha E_c \varphi \quad , \tag{5.14d}$$

$$E_\alpha \overline{\chi}_{\dot\beta} = -\tfrac{1}{24}(\gamma^{cde} P_\uparrow)_{\alpha\dot\beta} G_{cde} + \tfrac{1}{2}(\gamma^c P_\uparrow)_{\alpha\dot\beta} E_c \Phi - (\gamma^c \overline{\chi})_\alpha \overline{\chi}_{\dot\beta} E_c\varphi + \tfrac{1}{4} C_\alpha{}^{cd}(\gamma_{cd}\overline{\chi})_{\dot\beta} \quad , \tag{5.14e}$$

$$G_{\alpha\beta\gamma} = 0 \quad , \quad G_{\alpha bc} = 0 \quad , \tag{5.14f}$$

$$C_{\alpha b}{}^\gamma = -\tfrac{1}{8}(\gamma^{cd} P_\uparrow)_\alpha{}^\gamma (2C_{bcd} - C_{cdb} + G_{bcd}) \quad , \tag{5.14g}$$

$$E_\alpha C_{\beta c}{}^d = +\tfrac{1}{2} C_{\alpha\beta}{}^\epsilon C_{\epsilon c}{}^d + \tfrac{1}{2} C_{c(\alpha|}{}^e C_{e|\beta)}{}^d + \tfrac{1}{4}(\gamma^{ab})_{\alpha\beta}(G_{ca}{}^d - C_{ca}{}^d + C^d{}_{ca} + C^d{}_{ac}) E_b\varphi \quad . \tag{5.14h}$$

To be consistent within our teleparallel superspace, we use $E_a\varphi$, $E_a\widetilde{\varphi}$ instead of $\nabla_a\varphi$, $\nabla_a\widetilde{\varphi}$. However, they are actually the same, so that (5.9) is replaced by

$$E_{\underline{\alpha}}\varphi = E_{\underline{\alpha}}\widetilde{\varphi} = 0 \quad , \quad (E_a\varphi)^2 = (E_a\widetilde{\varphi})^2 = 0 \quad , \quad (E_a\varphi)(E^a\widetilde{\varphi}) = 1 \quad , \tag{5.15a}$$

$$E_a(E_b\varphi) = E_a(E_b\widetilde{\varphi}) = 0 \quad . \tag{5.15b}$$

Compared with the *non*-teleparallel case (5.8), the last terms in (5.14c) and (5.14e) are new ones, coming from the substitutions of (5.13a) for the Lorentz connection in the *non*-teleparallel equations. Eq. (5.14g) is also new here, because $T_{\alpha b}{}^{\gamma} = 0$ in (5.8f), but generated by the substitution of (5.13b). Eq. (5.14h) is another new one needed to satisfy the $(\underline{\alpha}\underline{\beta}c, d)$ and $(\alpha\beta\gamma, \delta)$-type Bianchi identities at $d = 1$. This is also because the superfield $C_{\alpha b}{}^c$ is treated as an independent superfield in our teleparallel superspace, similarly to the 11D case in (3.1e).

The extra constraints in our teleparallel superspace, corresponding to (5.10) are

$$C_{AB}{}^c E_c \varphi = 0 \ , \quad G_{ABc} E^c \varphi = 0 \ , \quad C_{aB}{}^C E^a \varphi = 0 \ , \tag{5.16a}$$

$$(E^a \varphi) E_a \Phi = 0 \ , \quad (E^a \varphi) E_a \overline{\chi}_{\dot{\beta}} = 0 \ , \tag{5.16b}$$

$$(\gamma^c)_\alpha{}^{\dot{\beta}} \overline{\chi}_{\dot{\beta}} E_c \widetilde{\varphi} = 0 \ , \quad C_{ab}{}^\gamma (\gamma^d)_\gamma{}^{\dot{\alpha}} E_d \widetilde{\varphi} = 0 \ . \tag{5.16c}$$

Needless to say, there is no constraints involving the curvature tensor or the Lorentz connection, such as (5.10b) or (5.10e).

Our constraints above can be derived from the *non*-teleparallel system [6][8], but we can independently confirm the satisfaction of Bianchi identities (5.11) and (5.12) at $0 \leq d \leq 1$. In this connection, useful relationships are

$$C_{\alpha b}{}^c E_c \widetilde{\varphi} = 0 \ , \tag{5.17a}$$

$$(C_{abc} - C_{acb} - C_{bca} + G_{abc}) E^c \widetilde{\varphi} = 0 \ , \tag{5.17b}$$

which are needed for confirming the $(ab\underline{\alpha}\underline{\beta})$ and $(\underline{\alpha}\underline{\beta}c, d)$-type Bianchi identities at $d = 1$. Note that $E_a \widetilde{\varphi}$ is used here instead of $E_a \varphi$. These constraints are not additional ones, but are just necessary conditions of (5.13).

We can mimic this procedure in other supergravity formulations in higher-dimensions in $D \geq 12$ [6][7][8], but we skip their details in this paper.

### 6. Concluding Remarks

In this paper we have presented a new formulation of superspace in which local Lorentz covariance is not manifest. We have set up all the necessary superspace constraints, and derived all the superfield equations as a result of satisfaction of all the Bianchi identities. There seems to be no fundamental obstruction for such a formulation in superspace.

The possibility of supergravity formulations in component that lack manifest local Lorentz covariance has been already known for some time. For example, in subsection 1.5 in [1], a component formulation using only torsions with no use of curvature tensors is mentioned as 'flat supergravity with torsion'. In this Letter, we have confirmed similar formulation is also possible in superspace for $D = 11$, $N = 1$ supergravity.

The important ingredients we have found in our work are summarized as follows: First, we have found that all the superspace constraints and superfield equations are re-formulated completely in terms of anholonomy coefficients $C_{AB}{}^C$ with no explicit usage of the Lorentz connection, as a parallel result to component formulation [1]. Second, we have found that the particular anholonomy coefficient component $C_{\alpha b}{}^c$ is to be treated as an independent superfield, because all the Bianchi identities are satisfied for any arbitrary form of $C_{\alpha b}{}^c$. Third, this particular component is eventually combined with $C_{abc}$ to form locally Lorentz covariant terms, such as the Ricci tensor combination (2.7), in our superfield equations. Fourth, we have found that this teleparallel superspace has no problem with the fermionic $\kappa$-symmetry of supermembranes, or it has even simpler and more natural results. Fifth, the teleparallel formulation is possible not only in 11D, but also in other dimensions. As an explicit example, we have applied this formulation to $D = 12$, $N = 1$ supergravity [6][8].

Our teleparallel superspace formulation will be of great importance, when considering supergravities in higher-dimensions given in [6][7][8], in which Lorentz covariance is not built-in from the outset. As an illustrative example, we have re-formulated $D = 12$, $N = 1$ superspace supergravity [6][8] in terms of teleparallel superspace. We have found there are 'bridge' relationships (5.13), that give important links between the original *non*-teleparallel superspace [6][8] and our present teleparallel superspace. It is also important that the second terms in (5.12) will not enter the combination $\phi_{Ad}{}^e(\widetilde{\mathcal{M}}_e{}^d)_B{}^C$. Therefore the lack of closed forms for $\phi_{ab}{}^c$ or $\phi_{\alpha b}{}^c$ will *not* matter in practice to derive the set of constraints for teleparallel superspace. We expect that this basic structure is common to other higher-dimensional supergravities in $D \geq 12$ [7][8].

The fact that the fermionic $\kappa$-symmetry of supermembrane in the Green-Schwarz formulation is simplified in teleparallel superspace indicates something deeper in general supergravity backgrounds for extended objects. It seems that the local Lorentz symmetry is not crucial even in conventional supergravity, not to mention higher-dimensional ones in 12D [6][7] related to F-theory [12] or 13D [8] related to S-theory [13]. In other words, when we need to study non-perturbative nature of M-theory [10], F-theory [12], S-theory [13], or higher-dimensional theories, local Lorentz symmetry is not the dominating symmetry, like the case of the original superstring theory formulated in the light-cone gauge [18]. As a matter of fact, there are other examples of this kind for superstring physics, *e.g.*, in ref. [19], null-vectors for constraints in superspace for Green-Schwarz $\sigma$-model $\beta$-functions were introduced. It seems that the loss of Lorentz symmetry in the Green-Schwarz formulation is inherent in superstring theories, which were originally formulated in the light-cone gauge. It is not surprising that this feature of superstring theories pops up in M-theory [10], F-theory [12], S-theory [13], or other higher-dimensional and more fundamental theories. As some readers have already noticed, another important supporting fact is the recent development of noncommutative geometry [14] associated with D-branes, in which the constant tensor $\theta^{\mu\nu}$ breaks the Lorentz covariance. These recent developments give more than enough motivation to consider teleparallel supergravity formulation in superspace.

Our result of teleparallel superspace in this paper is directly applicable to other conventional supergravity theories in other dimensions. Now that we have at hand the explicit example of teleparallel superspace applied to *non*-teleparallel $D = 12$, $N = 1$ superspace [6][8] an application, it is much easier to repeat similar analyses in other supergravity theories in higher-dimensions in $D \geq 12$ [6][7][8] as well as lower-dimensions $D \leq 10$.

We are grateful to W. Siegel for helpful discussions.